\documentclass[twocolumn,preprintnumbers,amsmath,amssymb]{revtex4}
\usepackage{graphicx}
\usepackage{dcolumn}
\usepackage{bm}
\usepackage[latin1]{inputenc}
\newcommand{\comment}[1]{}
\newcommand\etal{\mbox{\textit{et al.}}}

\begin{document}
\setlength{\unitlength}{0.7\textwidth}
\preprint{}

\title{Spectral imbalance and the normalized dissipation rate of turbulence}


\author{W.J.T. Bos}
\author{L. Shao}
\author{J.-P. Bertoglio}

\affiliation{%
 LMFA, UMR CNRS 5509\\ Ecole Centrale de Lyon - Universit\'e Claude Bernard Lyon 1 - INSA de Lyon\\ 69134 Ecully Cedex, France}


\begin{abstract}
The normalized turbulent dissipation rate $C_\epsilon$ is studied in decaying and forced  turbulence by direct numerical simulations, large-eddy simulations and closure calculations. A large difference in the values of $C_\epsilon$ is observed for the two types of turbulence. This difference is found at moderate Reynolds number and it is shown that it persists at high Reynolds number, where the value of $C_\epsilon$ becomes independent of the Reynolds number, but is still not unique. This difference can be explained by the influence of the nonlinear cascade time that introduces a spectral disequilibrium for statistically non-stationary turbulence. Phenomenological analysis yields simple analytical models which satisfactorily reproduce the numerical results. These simple spectral models also reproduce and explain the increase of $C_\epsilon$ at low Reynolds number that is observed in the simulations. 
\end{abstract}

\maketitle

\section{Introduction}

The normalized dissipation rate $C_\epsilon$ is a quantity that has been extensively studied in the literature. $C_\epsilon$ is defined as:
\begin{eqnarray}\label{eq1}
C_\epsilon=\frac{\epsilon \mathcal L}{\mathcal U^3},
\end{eqnarray}
in which $\epsilon$ is the viscous dissipation of turbulent kinetic energy, $\mathcal L$ is the integral length scale and $\mathcal U$ is the root-mean-square velocity. Recently Burattini, Lavoie and Antonia \cite{Burattini} revisited the behavior of $C_\epsilon$ and in particular discussed its dependency with the Reynolds number. Considerable scatter was observed in the data of various experiments. The same scatter was already observed by Sreenivasan \cite{Sreeni84}. The Direct Numerical Simulations (DNS) of isotropic turbulence reported in reference [\onlinecite{Sreeni98}] and [\onlinecite{Burattini}] (see for a compilation Kaneda \etal \cite{Kaneda}) exhibited significantly less scatter at large Reynolds number, which seemed to corroborate the general belief that, at high Reynolds numbers, $C_\epsilon$ tends to a constant value, at least for isotropic turbulence. 

In the present paper we address the issue of the high Reynolds number universality of the value of $C_\epsilon$. 
We investigate the case of isotropic turbulence, conceptually the simplest type of turbulence. Isotropic turbulence, either decaying or maintained statistically stationary by injecting energy at the large scales, is often considered as being close to spectral equilibrium. This is indeed true for stationary turbulence, but not for decaying turbulence, and it will be shown in the paper that the discrepancies existing between the two cases are sufficient to significantly affect the value of  $C_\epsilon$, even in the limit of high Reynolds number. The fact that the asymptotic value of  $C_\epsilon$ will be found not to be unique, even in the simple case of isotropic turbulence, will therefore provide an illustration there is no universal value of $C_\epsilon$  at high Reynolds number.

We note that high Reynolds number results for freely decaying fully developed turbulence  are hard to obtain by DNS, because at short times the DNS results are to a large extent determined by the initial conditions, whereas at large times the Reynolds number has already considerably decreased by viscous dissipation. 
In this work, we will therefore use DNS exclusively to study the lower Reynolds numbers. The behavior at large Reynolds will be investigated by Large Eddy Simulation (LES) and spectral closure.  

We perform LES, DNS and closure calculations of forced and decaying isotropic turbulence and the results are presented in section \ref{Calc}. 
In section \ref{sec2a}, the influence of the forcing method, the initial conditions and the resolution of the computations on the value of $C_\epsilon$ is investigated. In \ref{sec2b} the results are presented, discussed and compared to previous studies. In section \ref{SecAnalysis}, simple phenomenological models are introduced. They are found to reproduce the results of the simulations. The simple phenomenological arguments on which they are built then provide a possible scenario and a way to explain the non-universality of $C_\epsilon$ as well as its Reynolds number dependency.

\section{Normalized dissipation for decaying and forced isotropic turbulence\label{Calc}}
\subsection{Method \label{sec2a}}

The calculations in this section are aimed at studying the Reynolds number dependency of $C_\epsilon$ and at addressing the question of a possible difference in the values of $C_\epsilon$ for forced and decaying turbulence. DNS, LES and closure calculations are performed. The code used for the DNS and LES computations is a classical pseudo-spectral code with fourth-order Runge-Kutta time integration scheme. LES are performed at a resolution of $64^3$ and DNS at resolutions of $128^3$ and $256^3$ grid points. Freely decaying and forced turbulence simulations are performed.

\begin{figure} 
\setlength{\unitlength}{1.\textwidth}
\includegraphics[width=0.35\unitlength,angle=270]{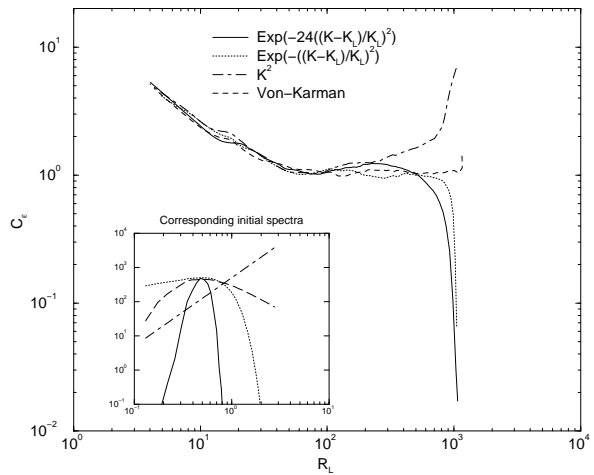}
\caption{ Evolution of $C_{\epsilon}$, starting from different initial
  energy spectra. The inset shows the initial spectral energy distributions $E(K)$  as a function of the wavenumber.\label{InitialSpecs} } 
\end{figure}

Before presenting and analyzing the results, it is necessary to investigate the influence of the initial conditions, the influence of the forcing scheme and finally the influence of the resolution and subgrid model. 
In the case of decaying turbulence, the influence of the initial conditions is checked by using four different initial spectra. The first and second initial spectra correspond to narrow spectral energy distributions:
\begin{eqnarray}\label{eqExpInit}
E(K,0)=exp\left(-24(K-K_L)^2/K_L^2\right)\nonumber\\
E(K,0)=exp\left(-(K-K_L)^2/K_L^2\right).
\end{eqnarray}
The third initial condition is an equipartition spectrum. In this case
the energy is equally distributed over all the wavevectors of the domain. The spherically averaged spectrum is then proportional to $K^2$. This is not a realistic energy distribution but it is used as a limiting case. The last initial spectrum is a von K\'arm\'an spectrum 
 \begin{equation}\label{Kar} 
E(K,0)=A\biggl({\frac{K}{K_{L}}}\biggr)^s\biggl(1 +\biggl({\frac{K}{K_{L}}}\biggr)^2\biggr)^{-(3s+5)/6}.
\end{equation} 
Note that this last spectrum behaves as $K^s$ at small $K$ and as $K^{-5/3}$ for large $K$. We use $s=4$.
The inset in figure \ref{InitialSpecs} shows the four different initial energy spectra.  
As illustrated in figure \ref{InitialSpecs}, after a transient period, the computations starting from all initial spectra yield values for $C_\epsilon$ that are nearly identical. A plateau around $1$ is found, regardless of the initial spectrum, and  all curves collapse for lower Reynolds number. The results corresponding to the exponentially decreasing initial spectra, show a transient that starts with a low value for $C_\epsilon$ because initially $\epsilon$ is small compared to the large scale quantity $\mathcal{U}^3\mathcal L^{-1}$. On the contrary, the equipartition distribution initially yields a large value for $C_\epsilon$, as the energy density at the high wavenumbers is initially very high. The von K\'arm\'an spectrum adapts fast to an equilibrium value, as the spectral distribution is initially close to a spectrum in equilibrium. These results show that the value of $C_\epsilon$ is independent of the initial conditions, once the turbulence has developed. In the following we will therefore limit the analysis to only one initial spectrum. We choose to  use the von K\'arm\'an spectrum (expression (\ref{Kar})), as it corresponds to the shortest transient.

\begin{figure} 
\setlength{\unitlength}{1.\textwidth}
\includegraphics[width=0.5\unitlength]{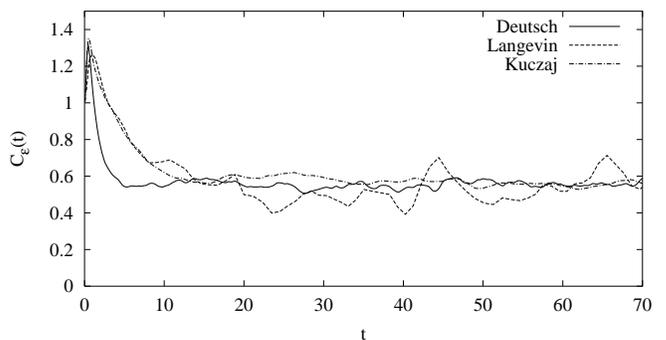}
\caption{Evolution of $C_{\epsilon}$ using different forcing schemes. \label{forcing} } 
\end{figure}

In the case of forced turbulence, a specific attention is paid to the way energy is injected in the simulation. 
Different forcing techniques are used in order to check the influence of the forcing schemes. 
The first one is the one originally proposed by Deutsch \cite{Deutsch}
that keeps the energy level in a low wavenumber band constant by
compensating what is removed by viscous and non-linear effects at each
time step.  In this technique, within a chosen spectral band $[K_1,K_2]$, the kinetic energy spectrum is therefore kept constant. An initial spectrum needs thus to be defined. The forcing spectrum used in this case is a von K\'arm\'an spectrum. The forced wavenumber band was varied from 2 to 5 wavenumbers of this spectrum and this  did not change significantly the value of $C_\epsilon$. Also using a $K^{-5/3}$ forcing spectrum did not significantly change the results. 

In the second technique a forcing term $f_i(\vec{K},t)$ is added to the Navier-Stokes equations. This force is defined as the solution of  a Langevin equation:
\begin{equation}
\frac{d}{dt}f_i(\vec{K},t)=-\frac{f_i(\vec{K},t)}{T}+g_i(\vec{K})~~~for~K_1~\le~|\vec{K}|~\le~K_2
\end{equation}
where ${T}$ is a forcing memory time and $g_i(\vec{K})$ is a white noise stochastic process with constant amplitude.

The last forcing scheme is:
\begin{equation}
f_i(\vec{K},t)=A_0\frac{u_i(\vec{K},t)}{|u(\vec{K},t)|^2} ~~~for~K_1~\le~|\vec{K}|~\le~K_2,
\end{equation}
with ${A_0}$ constant. Note that this is a particular case of the forcing used by Kuczaj \etal {\cite{Kuczaj} in their study of modulated turbulence. The resulting normalized dissipation rate, obtained with the three forcing schemes is shown in figure \ref{forcing} as a function of time normalized by the initial eddy turnover time. It can be observed that after a transient period all three methods yield values for $C_\epsilon(t)$ that oscillate around the same constant value. In the case of the forcing scheme determined by the Langevin equation, there is considerable oscillation of $C_\epsilon(t)$, but the mean value is approximately unaltered. 
The mean value of $C_\epsilon$ being unaffected by the type of forcing, we will use in the following section only one type, the first forcing technique (Deutsch \cite{Deutsch}). Another important conclusion that can be drawn from the results in figure \ref{forcing} is that if one wants to eliminate the effect of the instantaneous fluctuations on the value of $C_\epsilon$, averaging has to be applied over large time intervals (in some cases up to 10 initial eddy turnover times).

 \begin{figure} 
\setlength{\unitlength}{1.\textwidth}
\includegraphics[width=0.35\unitlength,angle=270]{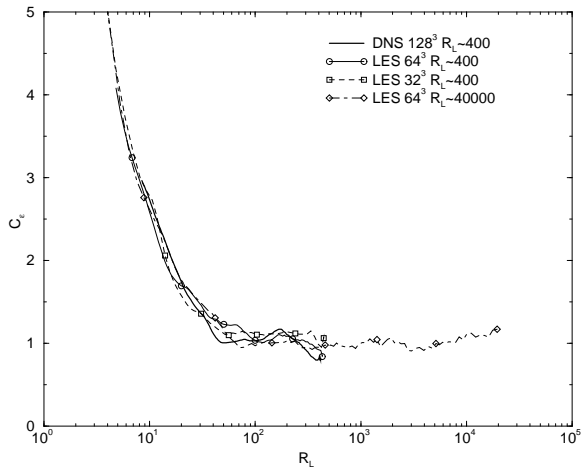}
\caption{$C_{\epsilon}$ as a function of the Reynolds number for decaying DNS and LES at different resolutions. The values of the Reynolds numbers indicated in the plot correspond to the initial Reynolds numbers of each simulation.\label{Resolutions} } 
\end{figure}

The subgrid model for the LES is the Chollet-Lesieur \cite{Chollet} eddy viscosity model, modified to account for finite-Reynolds-number effects (see references \onlinecite{Chollet83,Parpais}). The eddy viscosity in the Chollet-Lesieur model is
\begin{equation}\label{CLRE} 
\nu_t(K)=0.267\sqrt{\frac{E(K_c)}{K_c}}f\biggl(\frac{K_c}{K_{\eta}} \biggr)g\biggl(\frac{K}{K_c}\biggr) 
\end{equation} 
The function $g(\frac{K}{K_c})$ represents the `cusp' effect:
 \begin{equation} g\biggl(\frac{K}{K_c}\biggr)=1+1.7693\biggl(\frac{K}{K_c} \biggr)^{0.372} \end{equation} 
with $K_c$ the LES wavenumber cut-off. 
The function $f(K_c/K_{\eta})$, absent in the classical version of the model, corresponds to the low-Reynolds-number correction. It reads
 \begin{equation} 
f\biggl(\frac{K_c}{K_{\eta}}\biggr)=1-\left(\frac{K_c}{K_{\eta}}\right)^{4/3}\left(1+\frac{1}{a}ln\left(\frac{1+a\left(\frac{K_c}{K_{\eta}}\right)^{-4/3}}{1+a}\right)\right) 
\end{equation} 
 and $a$ is a constant equal to $0.38$. 
$K_{\eta}$ is the Kolmogorov wavenumber estimated using $K_{\eta }=(\epsilon /\nu ^3)^{1/4}$.  Use of (\ref{CLRE}) instead of the original Chollet-Lesieur \cite{Chollet} model has the advantage of permitting a smooth transition from LES to DNS as the Reynolds number decreases during the decay of the turbulence. 
This issue was already addressed in a previous work \cite{touil2} and good agreement with DNS was obtained as can also be observed in figure \ref{Resolutions}.
As a matter of fact, most of the decaying LES computations presented in the paper are indeed DNS at the end of their evolutions in time. In order to check the influence of the subgrid model, comparisons using the CZZS dynamic model \cite{Cui} were also performed at high Reynolds number. No significant influence on $C_\epsilon$ was found and consequently only the results obtained with the Chollet-Lesieur model are reported in the present paper.

Since the dissipative range of the spectrum is not, or not entirely, captured by LES, it is not possible to evaluate $\epsilon$ directly in this case. Instead, an estimate of $\epsilon$, $\epsilon^{LES}$, is evaluated, $\epsilon^{LES}$ being the sum of the energy flux to the small scales and the resolved part of the dissipation. It was checked by varying the resolution of the LES runs that the values of $C_\epsilon$ were not significantly altered by this approximation. In figure \ref{Resolutions} this is illustrated. The LES and DNS values collapse to one single curve for the different resolutions.

In addition to LES and DNS, closure calculations are performed. We use the EDQNM closure \cite{Orszag}. 
Although this closure was initially developed to study high Reynolds number turbulence, it has been shown to perform very well at low Reynolds numbers. In a recent work \cite{Bos2005} it was shown that the inertial range scaling exponents obtained by EDQNM compare very well to wind tunnel measurements of Mydlarski and Warhaft \cite{Mydlarski}. Furthermore, in the present paper it is shown that the EDQNM results are in good agreement with DNS and LES.
The exact formulation of the closure and the numerical method are the same as in Touil \etal \cite{touil2} and for details we refer to this work. The initial condition for the closure calculations is expression (\ref{Kar}), identical to the one  used in the DNS and LES. Also the large scale forcing is similar to the forcing adopted in DNS and LES ($E(K)$ being kept constant in the band $[K_1,K_m]$). The spectral resolution is approximately $20$ wavenumbers per decade.

\subsection{Results \label{sec2b}}

\begin{figure} 
\setlength{\unitlength}{1.\textwidth}
\includegraphics[width=0.5\unitlength]{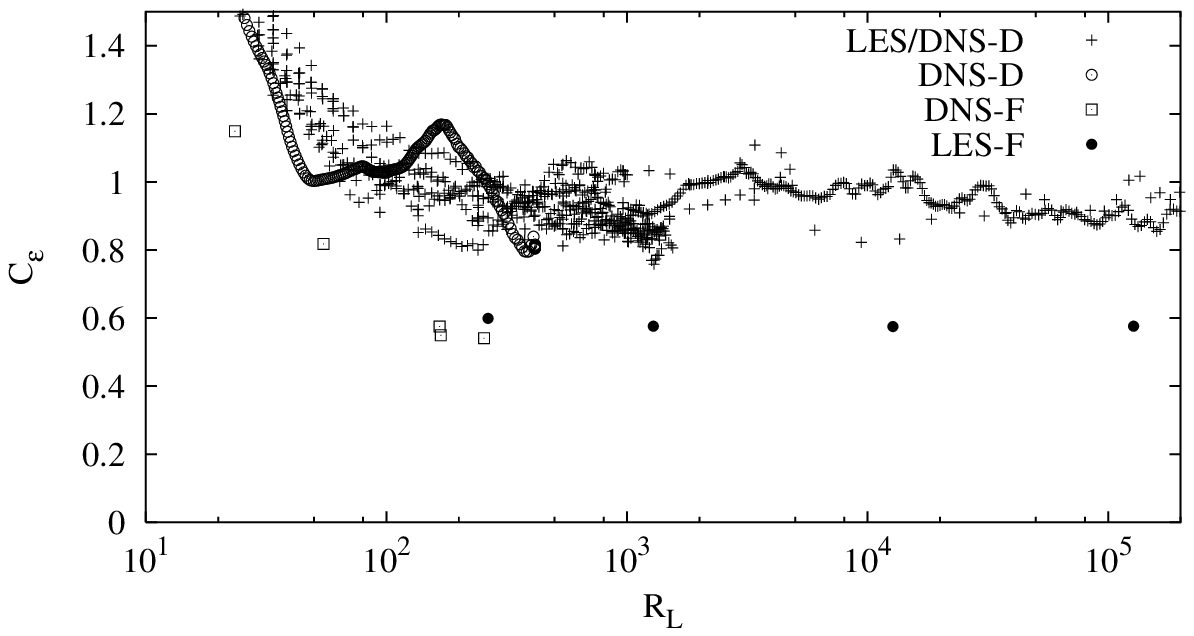}
\includegraphics[width=0.5\unitlength]{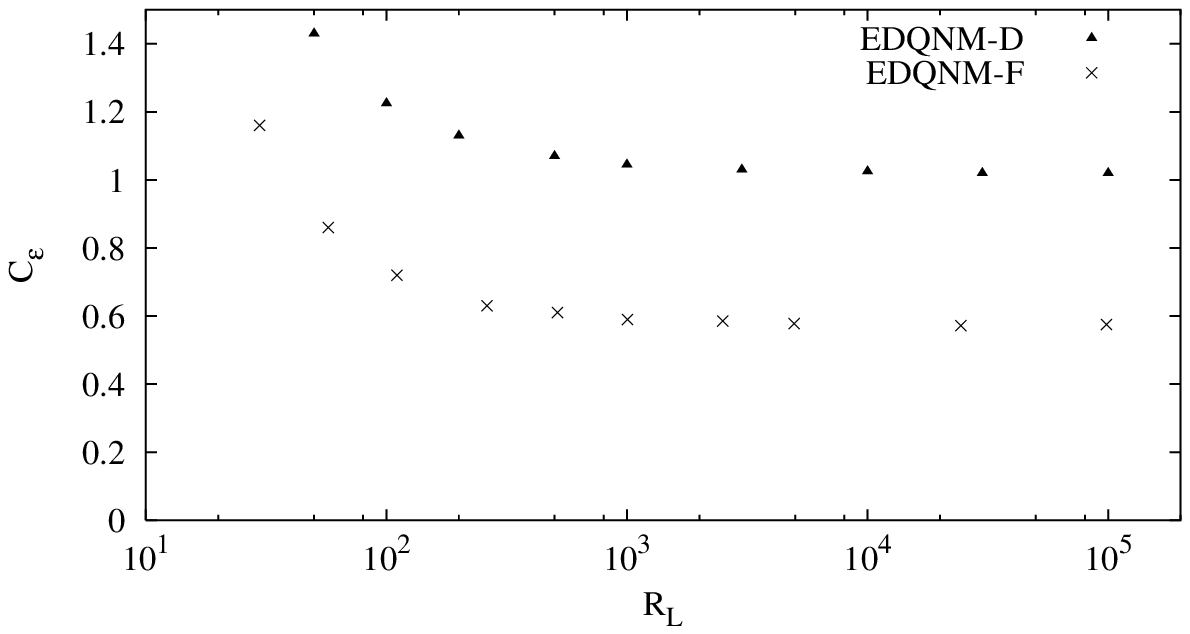}
\caption{Top: DNS and LES results for the normalized dissipation rate as a function of the Reynolds number. Bottom: Closure results. 
\label{ResLESDNSzoom}  }
\end{figure}

\begin{figure} 
\setlength{\unitlength}{1.\textwidth}
\includegraphics[width=0.5\unitlength]{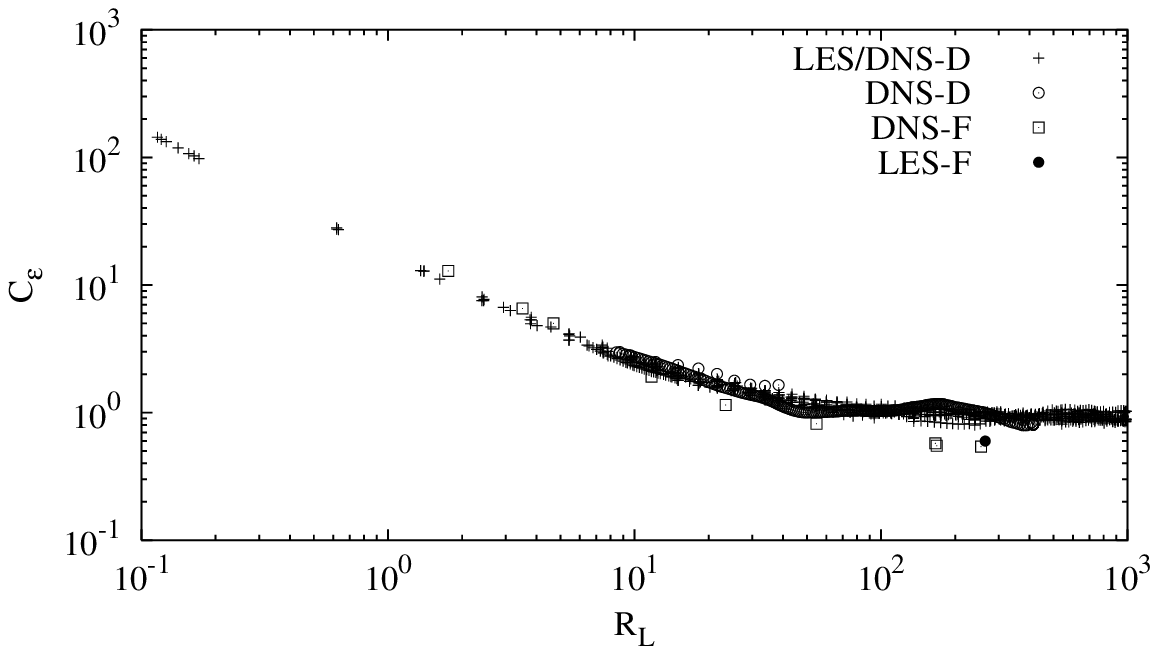}
\includegraphics[width=0.5\unitlength]{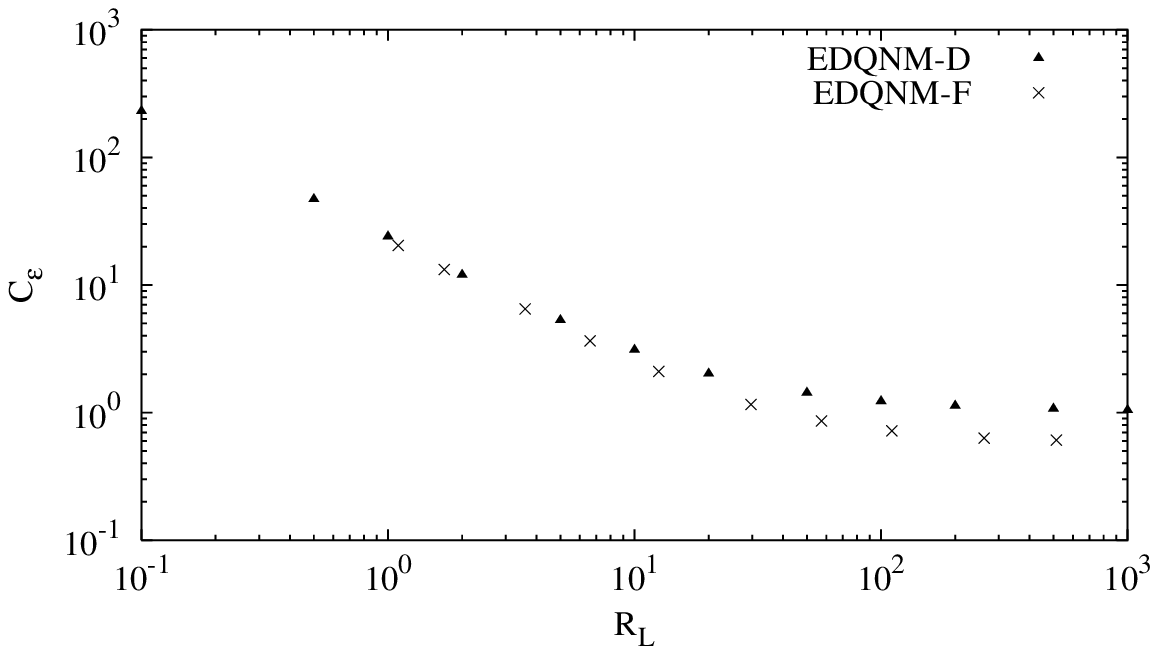}
\caption{ Same as figure \ref{ResLESDNSzoom}, but with a focus on the low Reynolds number range.\label{ResLESDNS}  }
\end{figure}

In figure \ref{ResLESDNSzoom}  the results of LES, DNS and closure are represented. $C_{\epsilon}$ is plotted as a function of the Reynolds number $R_L$, 
$R_L$ being directly accessible in the present calculations, including the case of the LES.
In figure \ref{ResLESDNSzoom} and the following figures, $D$ stands for decaying and $F$ for forced turbulence. The crosses correspond to LES results of decaying turbulence using the subgrid model with a Reynolds number correction which, as explained previously, allows for a smooth transition from LES to DNS. All results for Reynolds numbers smaller than $200$ can be considered to be DNS. At large $R_L$ all calculations of decaying turbulence tend to a value of approximately $1$. 

 The simulations of forced turbulence yield a value around $0.6$ for large Reynolds number. } The results for forced turbulence show less scatter because they correspond to numerically stationary situations and therefore they can be easily averaged. In Figure \ref{ResLESDNS} the results are shown for low Reynolds numbers.  For very small $R_L$, the difference between the values of $C_\epsilon$ for forced and decaying turbulence vanishes and all simulations yield results that tend to a $aR_L^{-1}$ dependence with $a\approx 20$. The results of the closure calculations show a very similar picture. There is no scatter in these results, because the closure approach is based on statistically averaged quantities and so does not take into account instantaneous turbulent fluctuations. 

\begin{figure} 
\setlength{\unitlength}{1.\textwidth}
\includegraphics[width=0.5\unitlength]{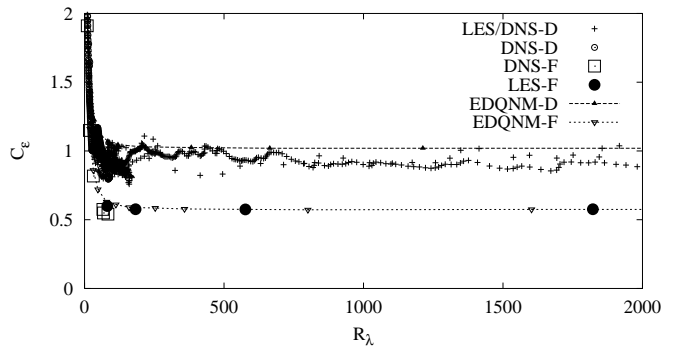}
\caption{DNS, LES and Closure results for the normalized dissipation rate as a function of the Taylor-scale Reynolds number.\label{ResLambda}  }
\end{figure}

In the present paper, $C_\epsilon$ is plotted as a function of $R_L=\mathcal{UL}/\nu$, the Reynolds number based on the integral length scale. 
In experiments, the integral length scale is not an easily accessible quantity, so that in most studies the Reynolds number $R_\lambda$, based on the Taylor micro-scale is used. The relation between  $R_L$ and $R_\lambda$ is 
\begin{eqnarray}
R_{\lambda}=\sqrt{\frac{15}{C_\epsilon}R_L}.
\end{eqnarray}
For comparison with other works on the subject, in figure \ref{ResLambda} the present results were replotted as a function of $R_\lambda$. Comparison of the results in this figure shows that, as noted before, at high Reynolds numbers, decaying turbulence and forced turbulence do not yield the same value for $C_\epsilon$, even though the value might slightly differ when changing the method.

At this point a discussion of the results of earlier studies on the value of $C_\epsilon$ is appropriate. First, in the case of Direct Numerical Simulations of statistically steady turbulence, most works are in overall agreement as can be seen for example in Kaneda \etal \cite{Kaneda}. The values vary between 0.41 and 0.69. This relatively high dispersion might be explained by the time over which the results are averaged, as noted in our discussion of the results in figure \ref{forcing}. For example, the study of Jim\'enez \etal \cite{Jimenez93} reports a value around $0.7$ for large Reynolds. However the largest Reynolds number calculation was run only for 0.3 turnover times, and it can be suspected that the results were influenced by time fluctuations. Furthermore the statistics were based on the time interval between 0.05 and 0.3 turnover times which might not be enough for the large scales to achieve a steady state. In the results of the DNS of decaying turbulence of Wang \etal \cite{Wang} the value  of $C_\epsilon$ (0.62 for $R_L=725$) is relatively low compared to the present results. 
However, only one value was reported for each decaying DNS simulation and it is therefore difficult to conclude about the value of $C_\epsilon$ by using their data. 

The agreement with experimental works is also difficult to assess. The results for decaying turbulence reported in Sreenivasan \cite{Sreeni84} seem to be in agreement with our conclusion, as are the results of Mydlarski and Warhaft \cite{Mydlarski3} who found in their decaying grid-generated turbulence a value of $0.9$. However it should be noted that the dissipation and integral length scale in experiments are difficultly accessible\cite{Burattini} and are not always evaluated in a uniform way. The experimental data by Pearson \etal \cite{Pearson} are perhaps most in disagreement with the present study. A relatively low value for $C_\epsilon$ is found for decaying grid turbulence. The reason for this needs further investigation. A possible explanation could be the presence of regions with shear or inhomogeneities due to the finite domain size.

This discussion stresses the importance of the present study in which decaying and forced turbulence are investigated using the same numerical methods, in which the different quantities are uniquely defined, evaluated in the same manner, and in which the effects of initial conditions and statistical averaging are carefully accounted for.

\section{Analysis \label{SecAnalysis}}


\begin{figure} 
\setlength{\unitlength}{1.\textwidth}
\includegraphics[width=0.45\unitlength]{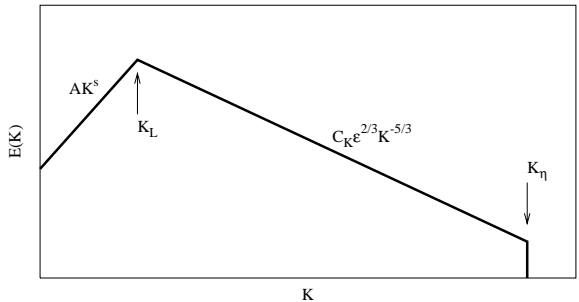}
\caption{ Model spectrum.\label{FigTP}}
\end{figure}

The aim of this section is to propose a possible scenario which explains the observation in the previous section that $C_\epsilon$ is larger in decaying turbulence than in forced turbulence using a simple spectral analysis.
The analysis will lead to analytical expressions for $C_\epsilon$ as a function of the Reynolds number. As a starting point we use the simple model spectrum (c.f. Comte-Bellot and Corrsin \cite{CBC}),
\begin{equation} 
E(K) = 
\begin{cases}\label{modspec}
AK^s & \textrm{for}~  K<K_{L}\\
C_K\epsilon^{2/3}K^{-5/3} & \textrm{for}~  K_L\le K \le K_\eta \\
0 & \textrm{for}~  K > K_\eta
\end{cases}
\end{equation}
where $C_K$ is the Kolmogorov constant. 
$A$ is assumed to be constant which can be considered valid for $s<4$, and a reasonable assumption for $s=4$ (see Lesieur and Schertzer \cite{Lesieur78}).

This spectrum corresponds to the sketch in figure \ref{FigTP}. Taking into account a more realistic shape of the turbulent spectrum, accounting for the rounding of the spectra around $K_L$ or a more refined dissipation range, does not considerably change the following results but yields rather complicated expressions. For simplicity's sake we use the simple form (\ref{modspec}).  
 
The kinetic energy, dissipation, root-mean-square velocity  and integral length scale are defined by:
\begin{eqnarray}\label{quants}
k=\int_0^{\infty}E(K)dK\nonumber\\
\epsilon=2\nu \int_0^{\infty}K^2E(K)dK\nonumber\\
\mathcal{U}=\sqrt{\frac{2}{3}k}\nonumber\\
\mathcal{L}=\frac{\pi}{2\mathcal{U}^2}\int_0^{\infty}K^{-1}E(K)dK
\end{eqnarray}
Assuming power law decay for the kinetic energy and dissipation:
\begin{eqnarray}\label{decaylaws}
k \sim t^{-n} \qquad&\qquad \epsilon \sim  nt^{-(n+1)}
\end{eqnarray}
and using the relation
\begin{eqnarray}
k_{,t} = -\epsilon 
\end{eqnarray}
one can, by substituting the model spectrum (\ref{modspec}) in the expressions (\ref{quants}), derive the usual decay exponent for isotropic turbulence: 
\begin{eqnarray}\label{ns}
n=\frac{2(s+1)}{s+3}.
\end{eqnarray}
For the integral length scale and timescale, defined as $\mathcal T =\mathcal {LU}^{-1}$, it is found
\begin{eqnarray}\label{tpropT}
\mathcal L \sim t^{1-n/2}\nonumber\\
\mathcal T\sim t^1.
\end{eqnarray}


\subsection{Influence of the cascade time}

We will now analyze why the value of $C_\epsilon$ is different for the cases of decaying and forced turbulence. This point can be made clear by considering the schematic picture of the 'perfect' energy cascade, which is energy conserving in between the large scales and the dissipation range, i.e. no energy is added or dissipated in between the beginning of the cascade at wavenumber $K_L$ and the exit of the wavenumber at $K_\eta$.  We introduce a cascade time corresponding to the time it takes for an amount of energy initially at $K_L$ to reach the dissipative scale $K_\eta=\beta K_L$. Following a reasoning by Lumley \cite{Lumley92}, stating that the local energy transfer at a wavenumber $K$ is entirely determined by $K$ and the local spectral energy density (see also Pope \cite{Pope}) this time $\mathcal T_c$ will be proportional to the integral time $\mathcal T$:
\begin{eqnarray}\label{Tc}
\mathcal T_c\sim \mathcal T (1-\beta^{-2/3})    
\end{eqnarray}

The rate at which energy is leaving the large scales and entering the cascade at time $t$ is governed by the large scale dynamics. It is therefore assumed to be:
\begin{equation}\label{epsfeps}
\epsilon_f(t)=C_\epsilon^f \frac{\mathcal U_{(t)}^3}{\mathcal L_{(t)}}
\end{equation}
where $\epsilon_f$ has not to be viewed as a dissipation, but as an energy flux, evaluated at the beginning of the cascade.  Note that for statistically stationary turbulence, the flux is constant in time so that 
\begin{equation}\label{epseps}
\epsilon_f(t)=\epsilon(t) 
\end{equation}
and consequently that $C_\epsilon^f=C_\epsilon^F$, where $C_\epsilon^F$ is the value of $C_\epsilon$ for forced turbulence, whose expression will be given in the next section. 

In the case of decaying turbulence (\ref{epseps}) is not holding, as the energy entering the cascade at time $t$ is not dissipated instantaneously. It only will be once it has reached the dissipative range of the spectrum, at time $t+\mathcal T_c$. Therefore instead of (\ref{epseps}) one has to write:
 \begin{equation}\label{epseps2}
\epsilon_f(t)=\epsilon(t+\mathcal T_c) 
\end{equation}
which together with (\ref{epsfeps}) yields:
\begin{equation}\label{epsfeps2}
\epsilon(t+\mathcal T_c)=C_\epsilon^F \frac{\mathcal U_{(t)}^3}{\mathcal L_{(t)}}.
\end{equation}
It is important to point out that since $\mathcal T_c$ is proportional to $t$ (expressions (\ref{tpropT}) and (\ref{Tc})), $\epsilon(t+\mathcal T_c)$ will still follow a power law decay proportional to $t^{-(n+1)}$. The presence of the cascade time will only influence the prefactor in the decay law. The introduction of a cascade delay time remains therefore compatible with the concept of self-similar decay.

Using the decay laws (\ref{decaylaws}) and (\ref{tpropT}) to express $\mathcal U$ and $\mathcal L$ in (\ref{epsfeps2}) at the same time as $\epsilon$ leads to:
\begin{equation}
\epsilon(t+\mathcal T_c)=C_\epsilon^F \frac{\mathcal U_{(t+\mathcal T_c)}^3}{\mathcal L_{(t+\mathcal T_c)}}\left(\frac{t}{t+\mathcal T_c}\right)^{-(n+1)}
\end{equation}
and therefore one immediately gets:
\begin{eqnarray}
\frac{C_\epsilon}{C_\epsilon^F}=\left(1+\frac{\mathcal T_c}{t}\right)^{n+1}.
\end{eqnarray}
 A finite cascade time yields then $C_\epsilon/C_\epsilon^F>1$, i.e. $C_\epsilon$ is not equal to and larger than $C_\epsilon^F$ for decaying turbulence.

Replacing $\mathcal T_c$ by its expression as a function of the integral time (\ref{Tc}), which during the decay process scales as $t$, leads to:
\begin{eqnarray}\label{CdCf}
\frac{C_\epsilon}{C_\epsilon^F}=\left(1+A_c(1-\beta^{-2/3})\right)^{n+1},
\end{eqnarray}
with $A_c$ a constant and $n$ defined in expression (\ref{ns}). The value of $A_c$ determines then $C_\epsilon/C_\epsilon^F$. For large $R_L$ expression (\ref{CdCf}) simplifies to:
\begin{eqnarray}\label{CdCf2}
\frac{C_\epsilon}{C_\epsilon^F}=\left(1+A_c\right)^n.
\end{eqnarray}
A value of $A_c=0.2$ agrees with  the value for $C_\epsilon/C_\epsilon^F$ observed in figure \ref{ResLESDNSzoom} at high $R_L$.

\subsection{Influence of the Reynolds Number}

After analyzing the effect of the turbulent cascade time in the case of non-stationary turbulence, we now address the problem of the influence of the Reynolds number on $C_\epsilon$, starting with the case of stationary turbulence. 
Stationary turbulence is a case where the spectrum is in strict equilibrium so that the value of the Kolmogorov constant is directly related to the value of $C_\epsilon$ as stated in Lumley \cite{Lumley92}. We will propose an analytical expression for $C_\epsilon$ as a function of the Reynolds number and of $s$, characterizing the large scales. This expression will then be coupled with the idea in the previous section.

We recall that we defined $\beta=K_\eta/K_L$. We will first consider the case $\beta>1$, corresponding to the presence of an inertial range. The case of smaller $\beta$, corresponding to the case where viscous effects are more important than nonlinear transfer will be discussed later. 

Calculating the relevant quantities defined in (\ref{quants}) using the model spectrum (\ref{modspec}), one obtains after some straightforward calculation:
\begin{eqnarray}\label{ceps1}
C_\epsilon^F=\frac{\pi \left(\frac{3s+5}{5s}-\frac{3}{5}\beta^{-5/3}\right)}
{2C_K^{3/2}\left(\frac{3s+5}{3(s+1)}-\beta^{-2/3}\right)^{5/2}}\\
R_L=\frac{\pi C_K^{3/2}\left(\frac{3s+5}{s}-3\beta^{-5/3} \right)  \left(3\beta^{4/3}-\frac{3s+5}{s+3}\right)          }
{20 \left(\frac{3s+5}{3(s+1)}-\beta^{-2/3}\right)^{1/2}}\label{RL1}
\end{eqnarray}
(\ref{ceps1}) and (\ref{RL1}) form a closed expression for the Reynolds number dependency of $C_\epsilon$ in stationary forced turbulence and depend on the parameters $s$, $\beta$ and the Kolmogorov constant $C_K$. In the case of large $\beta$, and thus large $R_L$, expression (\ref{ceps1}) simplifies to:
\begin{eqnarray}\label{cepsinf}
[C_\epsilon^F]_{\infty}=\frac{\pi (3(s+1))^{5/2}}
{10 C_K^{3/2} s(3s+5)^{3/2}}
\end{eqnarray}
yielding with $C_K=1.5$ and $s=4$ the value $[C_\epsilon^F]_{\infty}=0.53$.  
As mentioned before, in stationary turbulence $C_\epsilon$ is directly related to $C_K$. In addition to predicting the asymptotic value for $C_\epsilon$, expression (\ref{ceps1}) also predicts the low Reynolds number behavior of $C_\epsilon$ and the influence of the large scale topology, by including the parameter $s$. The influence of $s$ is relatively small. Varying $s$ from 2 to $\infty$ results in a variation of $C_\epsilon$ from $0.57$ to $0.51$.

\begin{figure} 
\setlength{\unitlength}{1.\textwidth}
\includegraphics[width=0.5\unitlength]{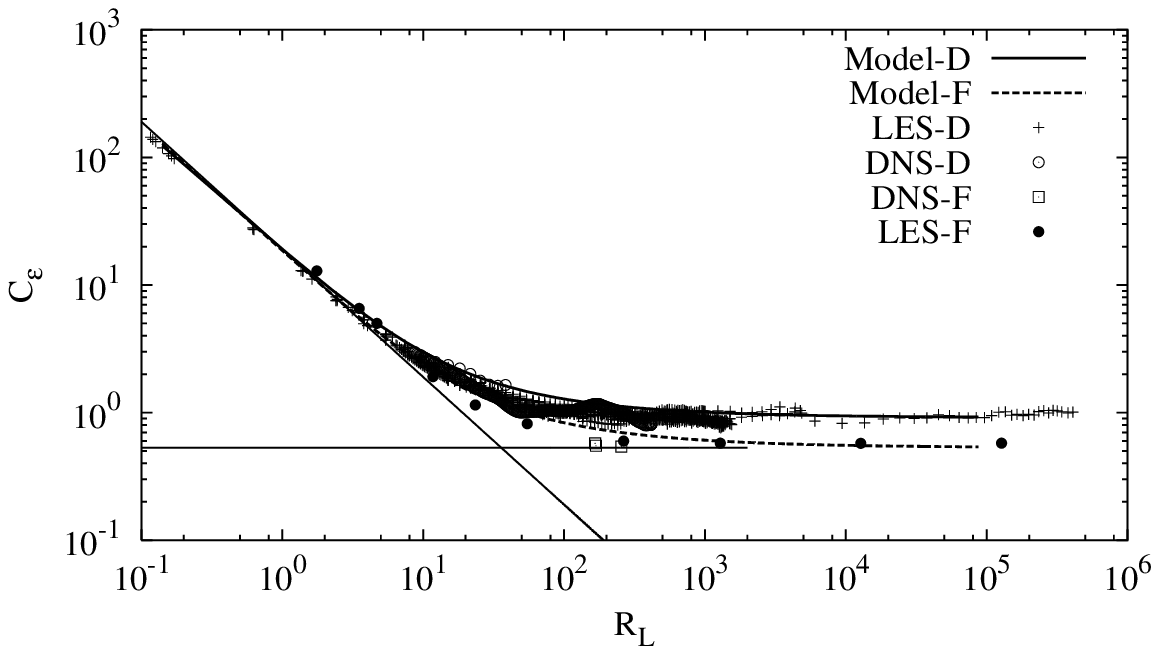}
\includegraphics[width=0.5\unitlength]{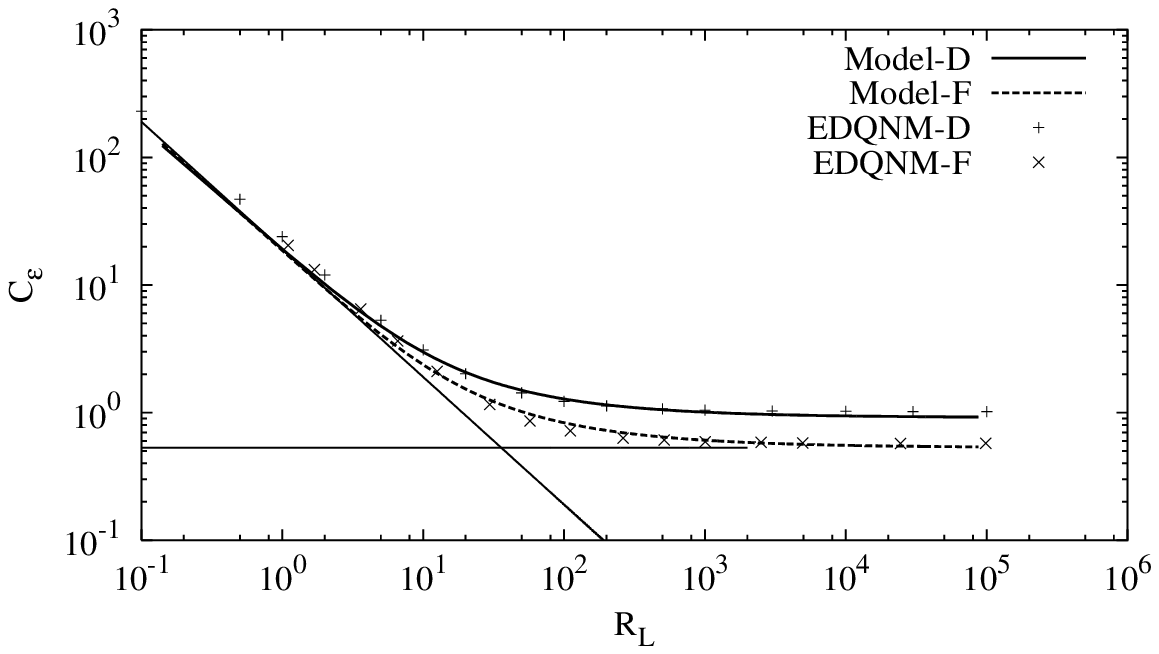}
\caption{ Top: comparison of DNS and LES results for the normalized
  dissipation rate with a simple phenomenological model. Bottom:
  comparison of closure results for the normalized dissipation rate
  with the same phenomenological model. The straight lines correspond
  to the low and high Reynolds number asymptotic limits, given by
  expressions (\ref{cepsinf}) and (\ref{cepsLow}).        \label{ModelDNSLES}}
\end{figure}

In figure \ref{ModelDNSLES} it is shown that both the asymptotic value and the full expression (\ref{ceps1}) describe the results  of the simulations of forced turbulence quite well. 

At low $R_L$ viscous effects become important even at the large scales. The eddy turnover time $\mathcal T \sim {\mathcal L}/{\mathcal U}$ is now not the only relevant timescale anymore at the large scales and so the value of $C_\epsilon$ changes. As can be seen in figure \ref{ResLESDNS}, this effect becomes noticeable at values of $R_L<10^3$. In the case of very low Reynolds number, the above analysis can not hold as there exists no inertial range. In this range we propose the analysis of a simplified spectrum:
\begin{equation} 
E(K) = 
\begin{cases}\label{modspecvis}
AK^s & \textrm{for}~  K<K_{\eta}\\
0 & \textrm{for}~  K > K_\eta
\end{cases}
\end{equation}
Note that this is a very rough representation of reality. A more realistic dissipation range will be considered below. Calculating $C_\epsilon$  using (\ref{modspecvis}) yields a $R_L^{-1}$ scaling, in agreement with Gebhardt \etal \cite{Gebhardt}. It is found:
\begin{eqnarray}\label{cepsLow}
[C_\epsilon^F]_{\nu}=\frac{27\pi^2 (s+1)^{3}}
{16 s^2(s+3)}R_L^{-1}.
\end{eqnarray}
For $s=4$ one finds $[C_\epsilon^F]_{\nu}\sim 19~R_L^{-1}$,\\
for $s=2$ one finds $[C_\epsilon^F]_{\nu}\sim 23~R_L^{-1}$. 

Expression (\ref{cepsLow}) is also shown in figure \ref{ModelDNSLES}.

To obtain an expression for the Reynolds number dependency of
$C_\epsilon$ in the case of decaying turbulence we multiply
(\ref{ceps1}) by the right hand side of (\ref{CdCf}) yielding:
\begin{eqnarray}\label{ceps1decay}
C_\epsilon=\frac{\pi \left(\frac{3s+5}{5s}-\frac{3}{5}\beta^{-5/3}\right)}
{2C_K^{3/2}\left(\frac{3s+5}{3(s+1)}-\beta^{-2/3}\right)^{5/2}}\left(1+A_c(1-\beta^{-2/3})\right)^n
\end{eqnarray}
In figure \ref{ModelDNSLES} it can be seen that the prediction for $C_\epsilon$ in decaying turbulence is also satisfactory. Taking into account the simplicity of the form of the model spectrum (\ref{modspec}), the agreement is fairly good.

\begin{figure} 
\setlength{\unitlength}{1.\textwidth}
\includegraphics[width=0.5\unitlength]{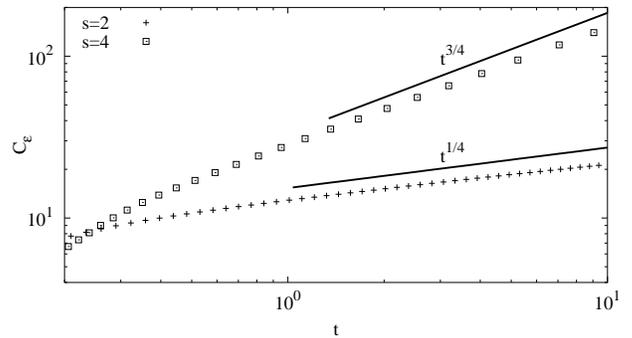}
\caption{Time dependence of the viscous decay of the normalized dissipation rate for initial spectra with a $K^2$ and $K^4$ small wavenumber range. \label{TimeViscous}}
\end{figure}

\begin{figure} 
\setlength{\unitlength}{1.\textwidth}
\includegraphics[width=0.5\unitlength]{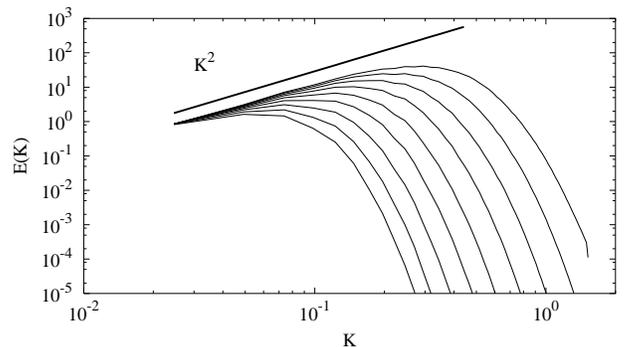}
\caption{Viscous decay of the energy spectrum with a $K^2$  small wavenumber range. \label{SpecViscous}}
\end{figure}

A more realistic study of the very low Reynolds number behavior can be proposed. It also provides a way to estimate the evolution of $C_\epsilon$ as a function of time. $C_\epsilon$ can be evaluated by assuming the time dependent spectrum,
\begin{eqnarray}
E(K,t)=E(K,0)e^{-2\nu K^2 t}  
\end{eqnarray}
and using $E(K,0)=AK^s$. After some straightforward calculation one finds:
\begin{eqnarray}
C_\epsilon^{\nu}(t)\sim t^{(s-1)/4}
\end{eqnarray}
In figure \ref{TimeViscous} this is checked by DNS for $s=2$ and $s=4$. The corresponding spectra are shown in figure \ref{SpecViscous} in the particular case where $s=2$. Note that for this more realistic dissipation range expressions are found for the asymptotically low Reynolds number behavior close to the ones obtained by using (\ref{cepsLow}). For $s=4$ one finds $[C_\epsilon^F]_{\nu}\sim 24~R_L^{-1}$ and for $s=2$ one finds $[C_\epsilon^F]_{\nu}\sim 32~R_L^{-1}$.

\section{Conclusion}

The behavior of the normalized dissipation rate $C_\epsilon$ was investigated in decaying and forced isotropic turbulence by DNS, LES and closure theory. It was shown that at moderate Reynolds numbers decaying turbulence yields a value for  $C_\epsilon$ different from, and higher than its value in forced turbulence. This difference persists at high Reynolds numbers and the value for $C_\epsilon$ in  decaying turbulence is about two times as high as its value in forced turbulence. It was shown by a simple phenomenological model that this difference is satisfactorily reproduced by introducing a finite cascade time to account for the imbalance between the large scales and the dissipation range. The Reynolds number dependency of $C_\epsilon$ was shown to be correctly predicted by analytically integrating a simple model spectrum. The time dependence of $C_\epsilon$ during the viscous decay was also estimated and agrees with results of low Reynolds number DNS.

An interesting direction for future work would be the investigation of other non-equilibrium situations such as shear flow in which the kinetic energy grows or rotating turbulence, in which the inertial range behavior is considerably modified \cite{LukasJoT}. Recent developments in subgrid modeling \cite{ShaoRot,LevequeShao} open perspectives to address these topics.


\end{document}